\documentclass[prl,aps,twocolumn,showpacs,preprintnumbers,amsmath,amssymb]{revtex4}
\usepackage{graphicx}

\begin{document}
\preprint{}
\draft

\title{Giant optical Faraday rotation induced by a single electron spin in a quantum dot: Applications to entangling remote spins via
a single photon}

\author{C.Y.~Hu}\email{chengyong.hu@bristol.ac.uk}
\author{A.~Young}
\author{J.L. O'Brien}
\author{J.G.~Rarity}
\affiliation{Department of Electrical and Electronic Engineering, University of Bristol, University Walk, Bristol BS8 1TR, United Kingdom}

\begin{abstract}
We propose a quantum non-demolition method - giant Faraday rotation - to detect a single electron spin in a quantum dot inside a
microcavity where negatively-charged exciton  strongly couples to the cavity mode.
Left- and right-circularly polarized light reflected from the cavity feels different phase shifts  due to cavity quantum electrodynamics
and the optical spin selection rule. This yields giant and tunable Faraday rotation which can be easily detected experimentally.
Based on this spin-detection technique, a scalable scheme to create an arbitrary amount of entanglement between two or more remote
spins via a single photon is proposed.
\end{abstract}

\date{\today}

\pacs{78.67.Hc, 03.67.Mn, 42.50.Pq, 78.20.Ek}

\maketitle

Photons and spins hold great potential in quantum information science, especially for  quantum communications, quantum information processing
and quantum networks \cite{nielsen00}. Photons are ideal candidates to transmit quantum information with little decoherence, whereas spins can
be used to store and process quantum information due to their long coherence times. Therefore investigations of spin manipulation, spin
detection, remote spin entanglement mediated by photons, and quantum state transfer between photons and spins are of great importance
\cite{loss98, imamoglu99, piermarocchi02, calarco03, yao05, clark07}.

Spin manipulation is well developed using pulsed magnetic resonance techniques, whereas single spin detection remains a challenging task.
Electrical detection of single spin has been reported in a gate-defined quantum box \cite{elzerman04, koppens06} and in a silicon field-effect
transistor \cite{xiao04}. The optically detected magnetic resonance technique (ODMR) proves to be an effective way to detect a single spin
either in a single molecule \cite{kohler93, wrachtrup93} or a single N-V center in diamond \cite{gruber97}.
 However, the ODMR technique  is based on the spin dependent fluorescence such that the spin is destroyed
after detection.  Recently, a non-demolition method to detect a single electron spin  has been experimentally reported by Berezovsky et al
\cite{berezovsky06} and Atat\"{u}re et al  \cite{atature07}. Both groups detect the tiny Faraday rotation angle induced by a single electron
spin in a quantum dot (QD), so the measured signals (even enhanced by a cavity) are rather weak and noisy.

It is widely accepted that entanglement is a useful  resource in quantum information science. Recently remote entanglement between photons,
trapped ions and atom ensembles have been demonstrated \cite{marcikic04, langer05, chou05}, however, all current experimental proposals for
entangling two atoms are restricted to one entanglement bit rather than  an arbitrary amount of entanglement \cite{lamata07, paternostro07}. To
our knowledge, entanglement between remote single spins has not yet been achieved  due to the lack of realizable proposals \cite{leuenberger05,
childress06, simon07}.

In this Letter, we propose a quantum  non-demolition method - giant Faraday rotation - to detect a single electron spin in a single QD inside a
microcavity. The different phase shifts for the left and right circularly polarized light reflected from the QD-cavity system yields giant
Faraday rotation which can be easily detected experimentally. This giant Faraday rotation induced by a single electron spin originates from the
spin dependent optical transitions of negatively-charged exciton and the effect of cavity quantum electrodynamics (cavity-QED). Based on this
spin detection technique, we propose a scalable  scheme to create an arbitrary amount of entanglement between two or more remote spins via a single photon, as well as the entanglement between single photons and single spins.

We consider a single self-assembled InGaAs/GaAs  QD inside a  micropillar cavity with circular cross section (see Fig. 1). This microcavity
consists of a $\lambda$-cavity between two GaAs/Al(Ga)As distributed Bragg reflectors. The QD is located in the center of the cavity to achieve
maximal light-matter coupling. This kind of structure as well as microdisks and photonic crystal nanocavities have been used to make single
photon sources \cite{moreau01, pelton02} and to study various cavity-QED effects \cite{reithmaier04, yoshie04, peter05}, such as the Purcell
effect in the weak-coupling regime, and the vacuum Rabi splitting in the strong coupling regime.

If the QD is neutral, optical excitation generates a neutral exciton. If the QD is singly charged, i.e., a single excess electron is injected,
optical excitation can create a negatively-charged exciton ($X^-$) which consists of  two electrons bound to one hole \cite{warburton97}. Due to
the Pauli's exclusion principle, $X^-$ shows spin-dependent optical transitions [see Fig. 1(b)] \cite{hu98}. If the excess electron lies in the
spin state $|+\frac{1}{2}\rangle \equiv |\uparrow\rangle$, only the left-handed circularly polarized light (marked by $|L\rangle$ or L-light)
can be resonantly absorbed to create  $X^-$ in the state $|\uparrow\downarrow\Uparrow\rangle$ with the two electron spins antiparallel. If the
excess electron lies in the spin state $|-\frac{1}{2}\rangle \equiv|\downarrow\rangle$, only the right-handed circularly polarized light (marked
by $|R\rangle$ or R-light) can be resonantly absorbed and create a $X^-$ in the state $|\uparrow\downarrow\Downarrow\rangle$. Here
$|\Uparrow\rangle$ and $|\Downarrow\rangle$ represent heavy-hole spin states $|\pm\frac{3}{2}\rangle$. Due to this spin selection rule, the L-
and R-light encounter different phase shifts after reflection from the $X^-$-cavity system, when $X^-$ strongly couples to the cavity. This will
be discussed below.

\begin{figure}[ht]
\centering
\includegraphics* [bb= 58 329 546 654, clip, width=6cm,height=5cm]{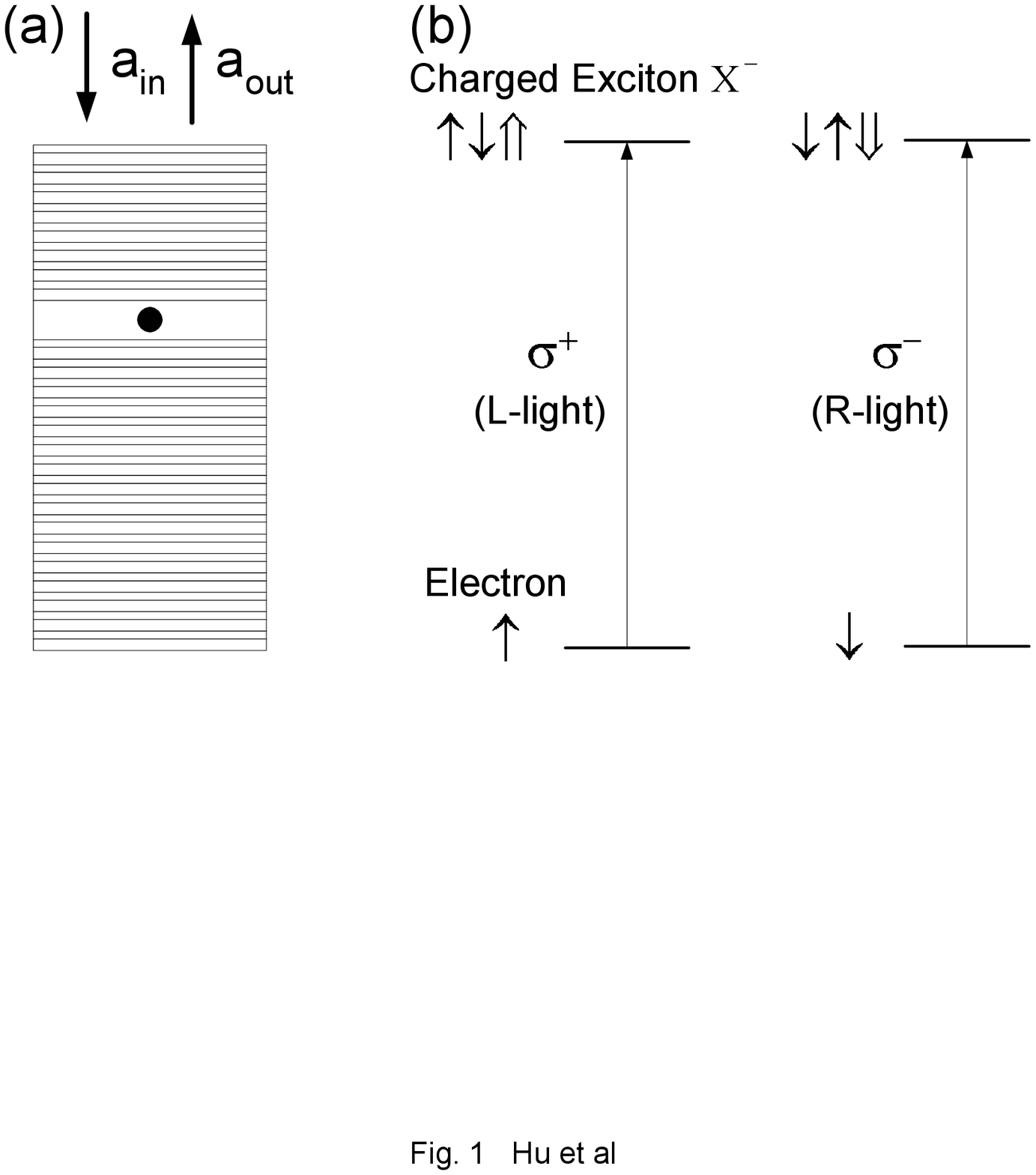}
\caption{(a) A charged QD inside a  micropillar microcavity with circular cross section. The distributed Bragg mirrors and the index guiding
provide three-dimensional confinement of light. (b) Spin selection rule for optical transitions of negatively-charged exciton $X^-$ in QD. If
the excess electron is in the spin state $|\uparrow\rangle$, only left-handed circularly polarized light (L-light) can create an $X^-$
transition. If the excess electron is in the spin state $|\downarrow\rangle$, only right-handed circularly polarized light (R-light) can create
an $X^-$ transition. This optical spin selection rule is due to the Pauli's exclusion principle.} \label{fig1}
\end{figure}

The Heisenberg equations for the cavity field operator $\hat{a}$ and QD dipole ($X^-$) operator $\sigma_-$ in the interaction picture, and the
input-output equation are given by \cite{walls94}
\begin{equation}
\begin{cases}
& \frac{d\hat{a}}{dt}=-[i(\omega_c-\omega)+\frac{\kappa}{2}]\hat{a}-\text{g}\sigma_--\sqrt{\kappa}\hat{a}_{in} \\
& \frac{d\sigma_-}{dt}=-[i(\omega_{X^-}-\omega)+\frac{\gamma}{2}]\sigma_--\text{g}\sigma_z\hat{a}+\hat{f} \\
& \hat{a}_{out}=\hat{a}_{in}+\sqrt{\kappa}\hat{a}
\end{cases}
\end{equation}
where $\omega$, $\omega_c$, $\omega_{X^-}$ are the frequencies of external field (probe beam), cavity mode, and $X^-$ transition, respectively.
g is the coupling constant between $X^-$ and the cavity mode. $\gamma/2$ is the QD dipole (($X^-$)) decay rate and $\kappa/2$ is the cavity field decay rate (side leakage from the cavity is neglected here). $\hat{f}$ is the noise operator needed to conserve the commutation relations.
$\hat{a}_{in}$ and $\hat{a}_{out}$ are the input and output field operators. For simplicity we consider the single-sided cavity with
the back mirror highly reflective and the front mirror partially reflective.

In the approximation of weak excitation, we can take $\langle \sigma_z\rangle \approx -1$. In the steady state, the reflection coefficient
for the QD-cavity system can be obtained
\begin{equation}
r(\omega)=1-\frac{\kappa[i(\omega_{X^-}-\omega)+\frac{\gamma}{2}]}{[i(\omega_{X^-}-\omega)+
\frac{\gamma}{2}][i(\omega_c-\omega)+\frac{\kappa}{2}]+\text{g}^2}.
\end{equation}
By taking $\text{g}=0$, we get the reflection coefficient for an empty cavity (cold cavity) with QD uncoupled to the cavity
\begin{equation}
r_0(\omega)=\frac{i(\omega_c-\omega)-\frac{\kappa}{2}}{i(\omega_c-\omega)+\frac{\kappa}{2}}
\end{equation}

\begin{figure}[ht]
\centering
\includegraphics* [bb= 24 327 558 724, clip, width=7cm,height=5.5cm]{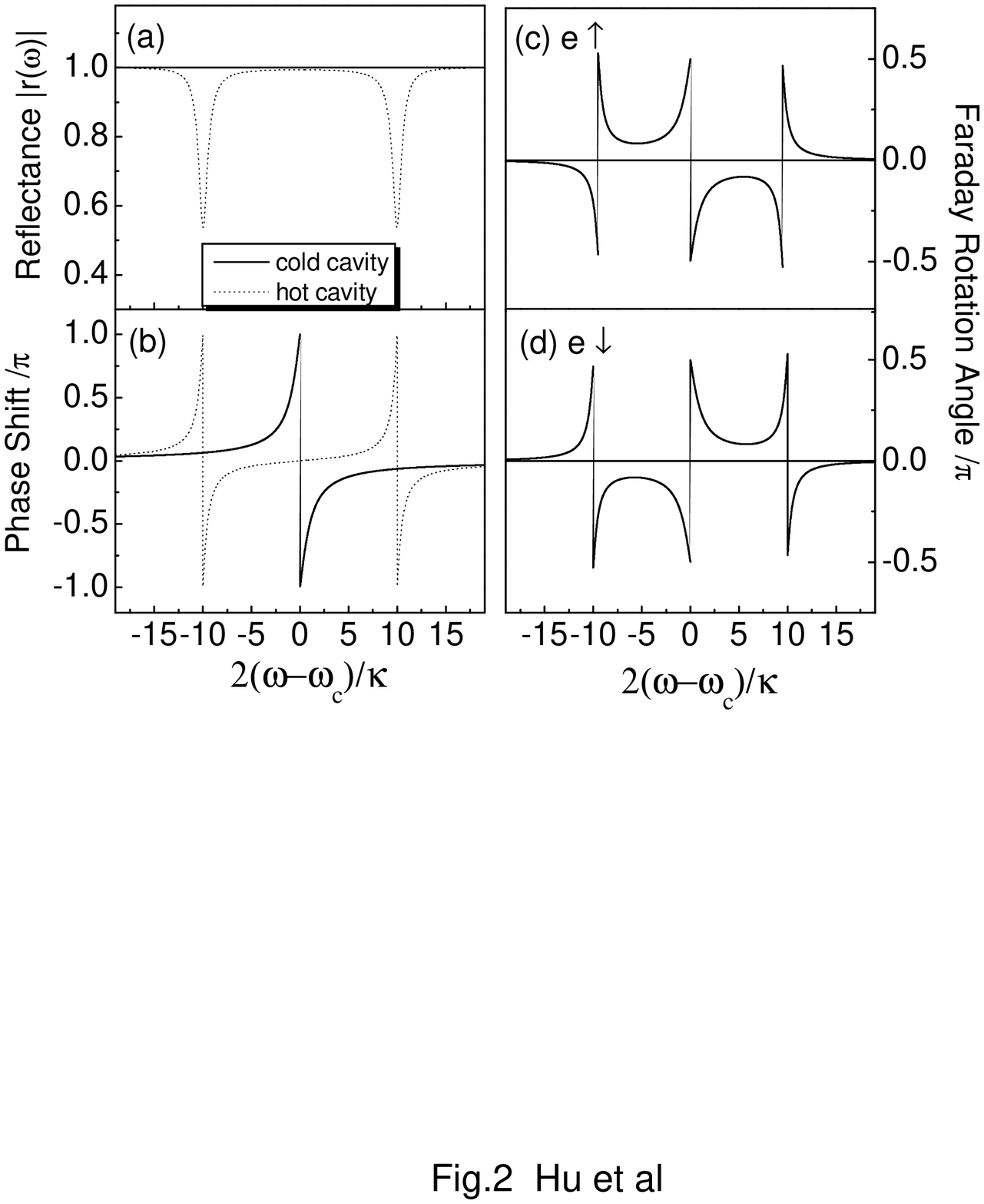}
\caption{Calculated (a) reflectance $|r(\omega)|$ and (b) phase shift vs frequency detuning from a cold cavity (solid curves) and a hot  cavity
(dotted curves). (c) and (d) show calculated Faraday rotation angle vs frequency detuning with the electron spin in the up or down states. We
take $\text{g}/\kappa=5.0$, $\gamma/\kappa=0.3$, and $\omega_{X^-}=\omega_c$.} \label{fig2}
\end{figure}

The complex reflection coefficients indicate that the reflected light feels a phase shift. The phase shift as a function of the frequency
detuning $\omega-\omega_c$ is presented in Fig. 2. For a cold cavity, the  phase shift is $\pm\pi$  at $\omega=\omega_c$, and decreases to zero
with increasing frequency detuning [solid curve in Fig. 2(b)]. In the strong-coupling regime with $\text{g}\gg (\kappa, \gamma)$, the $X^-$
state and cavity mode are mixed to form two new states, i.e., the dressed states, which leads to the vacuum-Rabi splitting. This state mixing
results in a zero phase shift at $\omega=\omega_c$ and two phase structures corresponding to the two dressed states [dotted curve in Fig. 2(b)].
The strongly coupled $X^-$-cavity system is called a hot cavity hereafter. In the following we show that the different phase shifts induced by a
cold cavity and a hot cavity can result in a giant Faraday rotation dependent on the state of the electron spin. We work near the resonant
condition with $|\omega-\omega_c| \ll \text{g}$ so that $|r(\omega)|=1$ holds for both the cold and the hot cavity [see Fig. 2 (a)].

As mentioned above, if the excess electron lies in the spin state $|\uparrow\rangle$, the L-light feels a hot cavity and gets a phase shift of
$\varphi_h$ after reflection, whereas the R-light feels the cold cavity and gets a phase shift of $\varphi_0$. Linearly polarized light as the
probe beam can be regarded as  the superposition of two circularly polarized components, i.e., $|R\rangle+|L\rangle$ (the factor $1/\sqrt{2}$ is
neglected). The reflected light then becomes $e^{i\varphi_0}|R\rangle+e^{i\varphi_h}|L\rangle$. The polarization direction of the reflected
light rotates an angle $\theta_F^{\uparrow}=\frac{\varphi_0-\varphi_h}{2}$, which is the so-called Faraday rotation.

Conversely, if the excess electron lies in the spin state $|\downarrow\rangle$, the R-light feels a hot cavity and get a phase shift of
$\varphi_h$ after reflection, whereas the L-light feels the cold cavity and gets a phase shift of $\varphi_0$. We thus get a Faraday rotation
angle $\theta_F^{\downarrow}=\frac{\varphi_h-\varphi_0}{2}=-\theta_F^{\uparrow}$. The sign of Faraday rotation angle depends on the electron
spin state.

Fig. 2 (c) and (d) present the calculated Faraday rotation angle vs the frequency detuning.  The Faraday rotation angles lie in the range
between $-\pi/2$ and $\pi/2$, which are huge compared with the conventional Faraday rotation \cite{berezovsky06, atature07,sugita03}. We call
this phenomenon  giant Faraday rotation by a single electron spin,  which is the result of cavity-QED as discussed above. The giant Faraday
rotation is partly supported by the experimental work from Kimble's group \cite{turchette95}, and can be easily detected experimentally.

If the single excess electron lies in a superposition spin state $|\psi\rangle=\alpha |\uparrow\rangle+ \beta |\downarrow\rangle$, after
reflection  the light and spin states become entangled,
\begin{equation}
\begin{split}
&(|R\rangle+|L\rangle) \otimes (\alpha |\uparrow\rangle+ \beta |\downarrow\rangle) \rightarrow e^{i\varphi_0}\times \\
&\left\{\alpha [|R\rangle+e^{i(\varphi_h-\varphi_0)}|L\rangle]|\uparrow\rangle + \beta
[e^{i(\varphi_h-\varphi_0)}|R\rangle+|L\rangle]|\downarrow\rangle \right\}.
\end{split}
\label{transform1}
\end{equation}
The reflected light is now in a mixed state of two linearly-polarized  components with Faraday rotation angles at
$\theta_F=\pm\frac{\varphi_0-\varphi_h}{2}$. When setting  $\varphi_0-\varphi_h=\pi/2$ by adjusting the frequency detuning to $\omega-\omega_c=-
\kappa/2$ [see Fig. 2 (b)], the probability to observe Faraday rotation angle at $\theta_F^{\uparrow}=\pi/4$ is $|\alpha|^2$, and that at
$\theta_F^{\downarrow}=-\pi/4$ is $|\beta|^2$. We can define the polarization degree
\begin{equation}
P_F=\frac{I(\pi/4)-I(-\pi/4)}{I(\pi/4)+I(-\pi/4)}
=\frac{|\alpha|^2-|\beta|^2}{|\alpha|^2+|\beta|^2},  \label{polarization}
\end{equation}
where $I(\pi/4)$ and $I(-\pi/4)$ are light intensity for Faraday rotation angle at $\pi/4$ and $-\pi/4$, respectively. $P_F$ is exactly the
electron spin polarization degree and this relation holds too if the electron spin is in a mixed state.  A proposed set-up to measure $P_F$ is
shown in Fig. 3(a) where the two light components with Faraday rotation angles at $\theta_F=\pm 45^0$ are analyzed by a half wave plate and a polarizing beam splitter. As the electron spin is not destroyed after measurements, this non-demolition spin detection method could be used to monitor the coherent electron spin precession if a transverse magnetic field is applied.

Note that the observation of the giant Faraday rotation relies on the strong coupling between $X^-$ and the cavity mode. The strong coupling
between  neutral exciton and the cavity mode has been observed in various microcavity geometries \cite{reithmaier04, yoshie04, peter05}. As the
$X^-$ transition can have larger oscillator strength than a neutral exciton, the strong coupling regime could be easier to attain for an $X^-$
in the same microcavity \cite{rapaport00}. In addition to single spin detection, the giant Faraday rotation could be utilized to make tunable single
QD wave plates, which will be discussed  elsewhere.

Up to now, the probe beam could be any classical or non-classical light at weak intensities. When using  single photons as the probe beam, we
can create entanglement between two remote spins in two spatially separated QD-cavity systems as shown in Fig. 3(b). The two QD-cavity systems
both work in the strong coupling regime, i.e., $X^-$ strongly couples to the cavity mode. For simplicity, we work near the resonant condition
with $|\omega-\omega_c| \ll \text{g}$ and $|r(\omega)|=1$, so that  $\varphi_h=0$  and $-\pi\leq \varphi_0 \leq \pi$ [see Fig. 2 (b)]. Now we
can make  phase shift gates and introduce the phase shift operator \cite{exp1}
\begin{equation}
\hat{U}(\varphi)=e^{i\varphi(|L\rangle \langle L|\otimes |\uparrow \rangle\langle \uparrow| + |R\rangle\langle R|\otimes |\downarrow
\rangle\langle \downarrow|)}, \label{phase}
\end{equation}
where $\varphi=\varphi_h-\varphi_0 \simeq -\varphi_0 $. In the first QD-cavity system, the excess electron is prepared in the spin state
$|\psi_1\rangle=\alpha_1 |\uparrow\rangle_1+ \beta_1|\downarrow\rangle_1$ with the corresponding phase shift operator $\hat{U}(\varphi_1)$; In
the second QD-cavity system, the excess electron  is prepared in the  spin state $|\psi_2\rangle=\alpha_2 |\uparrow\rangle_2+
\beta_2|\downarrow\rangle_2$ with the phase shift operator $\hat{U}(\varphi_2)$. Both QD-cavity systems are assumed to have the same
$\omega_c=\omega_{X^-}$.

\begin{figure}[ht]
\centering
\includegraphics* [bb= 90 326 506 657, clip, width=6cm, height=5cm]{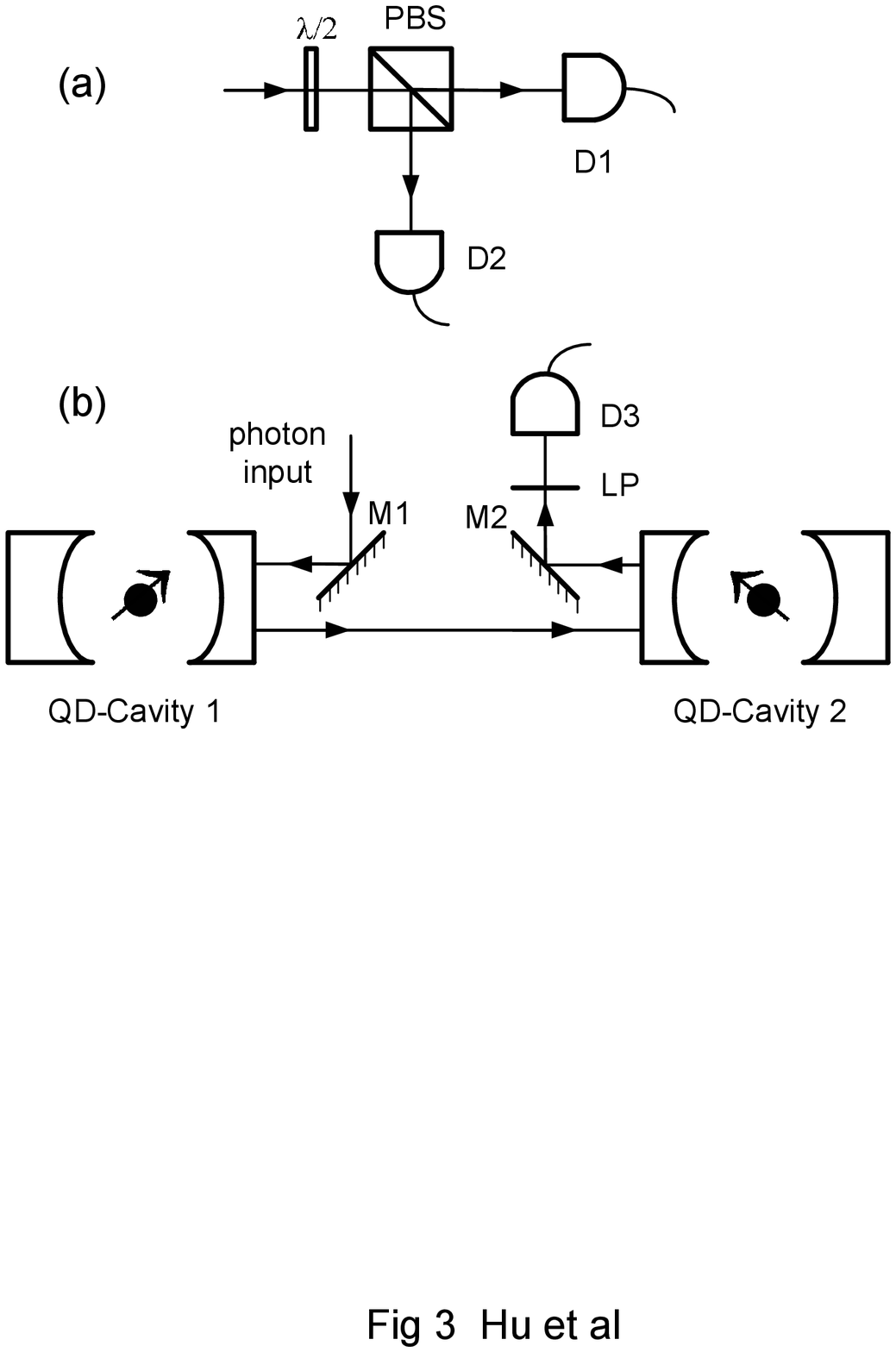}
\caption{(a) A proposed setup to measure the giant Faraday rotation and the electron spin polarization. $\lambda/2$ - wave plate, PBS -
polarized beam splitter, D1, D2 - detectors. (b) A proposed scheme to create entanglement between two remote spins via a single photon. Both QD-cavity
systems work in the strong-coupling regime. M1, M2 - reflection mirrors, LP - linear polarizer, D3 - detector.} \label{fig3}
\end{figure}

A linearly polarized single photon is reflected from the first cavity, then reflected from the second cavity, after which they are detected [see
Fig. 3(b)]. The corresponding state transformation is
\begin{equation}
\begin{split}
(|R\rangle & +|L\rangle)\otimes  (\alpha_1 |\uparrow\rangle_1+ \beta_1 |\downarrow\rangle_1)
\otimes (\alpha_2 |\uparrow\rangle_2+ \beta_2 |\downarrow\rangle_2) \\
~ \rightarrow  & ~ \alpha_1\alpha_2(|R\rangle+e^{i(\varphi_1+\varphi_2)}|L\rangle)|\uparrow\rangle_1|\uparrow\rangle_2 \\
& + \beta_1\beta_2(e^{i(\varphi_1+\varphi_2)}|R\rangle+|L\rangle)|\downarrow\rangle_1|\downarrow\rangle_2 \\
& + \alpha_1\beta_2(e^{i\varphi_2}|R\rangle+e^{i\varphi_1}|L\rangle )|\uparrow\rangle_1|\downarrow\rangle_2 \\
& + \beta_1\alpha_2(e^{i\varphi_1}|R\rangle+e^{i\varphi_2}|L\rangle)|\downarrow\rangle_1|\uparrow\rangle_2.
\end{split}
\label{transform2}
\end{equation}
When $\varphi_1=\varphi_2=\pi/2$ by adjusting the frequency detuning to $\omega-\omega_c=\kappa/2$ [see Fig. 2(b)], the output state becomes
\begin{equation}
\begin{split}
& ~ (|R\rangle-|L\rangle)\left[\alpha_1\alpha_2|\uparrow\rangle_1|\uparrow\rangle_2-
\beta_1\beta_2|\downarrow\rangle_1|\downarrow\rangle_2\right]\\
& +i(|R\rangle+|L\rangle)\left[\alpha_1\beta_2|\uparrow\rangle_1|\downarrow\rangle_2+\alpha_2\beta_1
|\downarrow\rangle_1|\uparrow\rangle_2\right].
\end{split}
\label{transform3}
\end{equation}
The output photon states can be measured in orthogonal linear polarizations. If detecting the photon in the $|R\rangle-|L\rangle$ state
($0^\circ$ linear), we project Eq. (\ref{transform3}) onto a two-spin entangled state
\begin{equation}
|\Phi_{12}\rangle=\alpha_1\alpha_2|\uparrow\rangle_1|\uparrow\rangle_2-\beta_1\beta_2|\downarrow\rangle_1|\downarrow\rangle_2.
\label{ent1}
\end{equation}
On detecting the photon in the $|R\rangle+|L\rangle$ state ($90^\circ$ linear), we project onto another two-spin entangled state
\begin{equation}
|\Psi_{12}\rangle=\alpha_1\beta_2|\uparrow\rangle_1|\downarrow\rangle_2+\alpha_2\beta_1 |\downarrow\rangle_1|\uparrow\rangle_2.
\label{ent2}
\end{equation}
Obviously, if the electron spins are prepared in arbitrary superposition states (which will be discussed later), we get an arbitrary amount of
entanglement between remote spins, which could be used as a resource for quantum communications and quantum information processing
\cite{lamata07}.

This scheme can be extended to create entanglement between three or more remote spins. For example, when using the  state in Eq.
(\ref{transform2}) as the input of the third QD-cavity with the  electron spin in the state $|\psi_3\rangle=\alpha_3 |\uparrow\rangle_3+
\beta_3|\downarrow\rangle_3$ and the phase shift operator $\hat{U}(\varphi_3)$, and setting $\varphi_1=\varphi_2=\varphi_3=\pi/2$, we can get a
three-spin entangled state
\begin{equation}
\begin{split}
|\Phi_{123}\rangle=& ~ \alpha_1\alpha_2\alpha_3|\uparrow\rangle_1|\uparrow\rangle_2|\uparrow\rangle_3-\beta_1\beta_2\alpha_3
|\downarrow\rangle_1|\downarrow\rangle_2|\uparrow\rangle_3 \\
&-\alpha_1\beta_2\beta_3|\uparrow\rangle_1|\downarrow\rangle_2|\downarrow\rangle_3 -\beta_1\alpha_2\beta_3
|\downarrow\rangle_1|\uparrow\rangle_2|\downarrow\rangle_3,
\end{split}
\end{equation}
when we detect the output photon  in the $|R\rangle-i|L\rangle$ state ($+45^\circ$ linear), and a similar state with $\alpha$'s replaced by
$\beta$'s and $|\uparrow\rangle$ by $|\downarrow\rangle$ when the output photon is detected in the $|R\rangle+i|L\rangle$ state ($-45^\circ$
linear). Another route to build higher-order entangled states and cluster states would be to sequentially entangle pairs of the spins by
repeating the above single-photon measurement scheme which leads to Eqs. (\ref{ent1}) -(\ref{ent2}), combined with controlled local
spin rotations.

Different from the proposals for entangling remote atoms based on interference of emitted photons \cite{langer05, chou05,
childress06, simon07}, our scheme utilizes a single-photon quantum bus to couple or entangle remote spins.
The  superposition state of the single electron spin in QD is a prerequisite to create entanglement between remote spins in our case.
This superposition state can be prepared, for example, by optical pumping and/or optical cooling \cite{berezovsky06, atature07,
atature06} using the side excitation, or by  electrical spin injection using techniques being developed in spintronics, therefore an arbitrary
amount of entanglement can be created.  Due to the long coherence time of electron spin in QD, the entanglement should be robust and  persist
for a long time ( $\sim \mu$s), so one can apply a magnetic field or microwave pulses to manipulate the entanglement. Finally, we can do quantum
state transfer between photons and spins with the spin-photon entanglement as demonstrated in Eq. (\ref{transform1}). Moreover, an arbitrary
amount of polarization entanglement between photons can be created either via a single spin followed by measuring the spin states, or via two
entangled remote spins followed by $X^-$ emissions \cite{hu07}.

In conclusion, giant optical Faraday rotation induced by a single electron spin in a QD is proposed as a result of cavity-QED. This enables us
to perform  non-demolition single spin detection and to build a tunable quantum phase  gate. Based on it, we have proposed a  scalable scheme to create an arbitrary amount of entanglement either between  remote spins or between  a single photon and
a single spin. This work opens a new avenue to build solid-state quantum networks with single photons and single QD spins.

J.G.R is supported by a Wolfson merit award.  This work is partly funded by UK EPSRC IRC in Quantum Information Processing and QAP (EU IST
015848).


\begin{references}

\bibitem{nielsen00}M.A. Nielson and I.L. Chuang, {\it Quantum Computation and Quantum Information}
(Cambridge Univ. Press, Cambridge, UK, 2000).

\bibitem{loss98}D. Loss and D.P. DiVincenzo, Phys. Rev. A {\bf 57}, 120 (1998).

\bibitem{imamoglu99}A. Imamoglu et al, Phys. Rev. Lett. {\bf 83}, 4204 (1999).

\bibitem{piermarocchi02}C. Piermarocchi et al, Phys. Rev. Lett. {\bf 89}, 167402 (2002).

\bibitem{calarco03}T. Calarco et al, Phys. Rev. A {\bf 68}, 012310 (2003).

\bibitem{yao05}W. Yao, R.-B. Liu, and L. J. Sham, Phys. Rev. Lett. {\bf 95}, 030504 (2005).

\bibitem{clark07}S.M. Clark et al, Phys. Rev. Lett. {\bf 99}, 040501 (2007).

\bibitem{elzerman04}J.M. Elzerman et al, Nature {\bf 430}, 431 (2004).

\bibitem{koppens06}F.H.L. Koppens et al,  Nature {\bf 442}, 766 (2006).

\bibitem{xiao04}M. Xiao et al, Nature {\bf 430}, 435 (2004).

\bibitem{kohler93}J. K\"{o}hler et al, Nature {\bf 363}, 242 (1993).

\bibitem{wrachtrup93}J. Wrachtrup et al, Nature {\bf 363}, 244 (1993).

\bibitem{gruber97}A. Gruber et al, Science {\bf 276}, 2012 (1997).

\bibitem{berezovsky06}J. Berezovsky et al, Science {\bf 314}, 1916 (2006).

\bibitem{atature07}M. Atat\"{u}re et al, Nature Physics {\bf 3}, 101 (2007).

\bibitem{marcikic04}I. Marcikic et al, Phys. Rev. Lett. {\bf 93}, 180502 (2004).

\bibitem{langer05}C. Langer et al, Phys. Rev. Lett. {\bf 95}, 060502 (2005).

\bibitem{chou05}C.W. Chou et al, Nature {\bf 438}, 828 (2005).

\bibitem{lamata07}L. Lamata, J.J. Garc\'{i}a-Ripoll, and J.I. Cirac, Phys. Rev. Lett. {\bf 98}, 010502 (2007).

\bibitem{paternostro07}M. Paternostro, M.S. Kim, and G.M. Palma, Phys. Rev. Lett. {\bf 98}, 140504 (2007).

\bibitem{leuenberger05}M.N. Leuenberger, M.E. Flatt\'{e} and D. D. Awschalom, Phys. Rev. Lett. {\bf 94}, 107401 (2005).

\bibitem{childress06}L. Childress et al, Phys. Rev. Lett. {\bf 96}, 070504 (2006).

\bibitem{simon07}C. Simon et al, Phys. Rev. B {\bf 75}, 081302(R) (2007).

\bibitem{moreau01}E. Moreau et al, Appl. Phys. Lett. {\bf 79}, 2865 (2001).

\bibitem{pelton02}M. Pelton et al, Phys. Rev. Lett. {\bf 89}, 233602 (2002).

\bibitem{reithmaier04}J.P. Reithmaier et al, Nature {\bf 432}, 197 (2004).

\bibitem{yoshie04}T. Yoshie et al,  Nature {\bf 432}, 200 (2004).

\bibitem{peter05}E. Peter et al, Phys. Rev. Lett. {\bf 95}, 067401 (2005).

\bibitem{warburton97}R.J. Warburton et al, Phys. Rev. Lett. {\bf 79}, 5282 (1997).

\bibitem{hu98}C.Y. Hu et al, Phys. Rev. B {\bf 58}, R1766 (1998).

\bibitem{walls94}D.F. Walls and G.J. Milburn, {\it Quantum Optics} (Springer-Verlag, Berlin, 1994).

\bibitem{exp0}The side leakage rate can be neglected if it is less than the cavity field decay rate.

\bibitem{sugita03}M. Sugita, S. Machida, and Y. Yamamoto,
Preprint at http://www.arxiv.org/quant-ph/0301064 (2003).

\bibitem{turchette95}Q.A. Turchette et al, Phys. Rev. Lett. {\bf 75}, 4710 (1995).

\bibitem{rapaport00}R. Rapaport et al, Phys. Rev. Lett. {\bf 84}, 1607 (2000).

\bibitem{exp1}The  fidelity for this phase gate increases with increasing the coupling strength $\text{g}$ for a perfect QD.
When $\text{g}\geq 1.5\kappa$, the fidelity is larger than 0.998.

\bibitem{atature06}M. Atat\"{u}re et al, Science {\bf 312}, 551 (2006).

\bibitem{hu07}C.Y. Hu et al, (unpublished).


\end{references}
\end{document}